\documentstyle[12pt,epsf,epsfig]{article} \hoffset -0.5in \textwidth 6.00in 
\textheight
8.5in  \parskip 7pt \openup1\jot \parindent=0.5in
\newskip\humongous \humongous=0pt plus 1000pt minus 1000pt
\def\caja{\mathsurround=0pt}
\def\eqalign#1{\,\vcenter{\openup1\jot \caja
        \ialign{\strut \hfil$\displaystyle{##}$&$
        \displaystyle{{}##}$\hfil\crcr#1\crcr}}\,}
\newif\ifdtup

\def\eqright #1\cr{\noalign{\hfill$\displaystyle{{}#1}$}}
\def\eqleft #1\cr{\noalign{\noindent$\displaystyle{{}#1}$\hfill}}
\def\oldreffmt#1{\rlap{[#1]} \hbox to 2\parindent{}}

\def\figfmt#1{\rlap{Figure {#1}} \hbox to 1in{}}

\def\sectioneq{\def\theequation{\thesection.\arabic{equation}}{\let
\holdsection=\section\def\section{\setcounter{equation}{0}\holdsection}}}%

\newcounter{holdequation}

\def\auto{\eqno(\refstepcounter{equation}\theequation)}
\def\begineq #1\endeq{$$ \refstepcounter{equation}\eqalign{#1}\eqno
	(\theequation) $$}
\def\contlimit{\,{\hbox{$\longrightarrow$}\kern-1.8em\lower1ex
\hbox{${\scriptstyle (a\rightarrow0)}$}}\,}
\def\centeron#1#2{{\setbox0=\hbox{#1}\setbox1=\hbox{#2}\ifdim
\wd1>\wd0\kern.5\wd1\kern-.5\wd0\fi
\copy0\kern-.5\wd0\kern-.5\wd1\copy1\ifdim\wd0>\wd1
\kern.5\wd0\kern-.5\wd1\fi}}
\def\centerover#1#2{\centeron{#1}{\setbox0=\hbox{#1}\setbox
1=\hbox{#2}\raise\ht0\hbox{\raise\dp1\hbox{\copy1}}}}
\def\centerunder#1#2{\centeron{#1}{\setbox0=\hbox{#1}\setbox
1=\hbox{#2}\lower\dp0\hbox{\lower\ht1\hbox{\copy1}}}}
\def\lsim{\;\centeron{\raise.35ex\hbox{$<$}}{\lower.65ex\hbox
{$\sim$}}\;}
\def\gsim{\;\centeron{\raise.35ex\hbox{$>$}}{\lower.65ex\hbox
{$\sim$}}\;}
\def\super#1{\ifmmode \hbox{\textsuper{#1}}\else\textsuper{#1}\fi}
\def\textsuper#1{\newcount\holdspacefactor\holdspacefactor=\spacefactor
$^{#1}$\spacefactor=\holdspacefactor}
\def\getcite#1,{\advance\citenumber by1
\def\getcitearg{#1}\def\lastarg{@}
\ifnum\citenumber=1
\ref{#1}\let\next=\getcite\else\ifx\getcitearg\lastarg\let\next=\relax
\else ,\ref{#1}\let\next=\getcite\fi\fi\next}
\def\pom{{\rm P\kern -0.53em\llap I\,}}
\def\spom{{\rm P\kern -0.36em\llap \small I\,}}
\def\sspom{{\rm P\kern -0.33em\llap \footnotesize I\,}}

\relax

\def\auto{\eqno(\refstepcounter{equation}\theequation)}
\def\begineq #1\endeq{$$ \refstepcounter{equation}\eqalign{#1}\eqno
	(\theequation) $$}
\def\contlimit{\,{\hbox{$\longrightarrow$}\kern-1.8em\lower1ex
\hbox{${\scriptstyle (a\rightarrow0)}$}}\,}
\def\centeron#1#2{{\setbox0=\hbox{#1}\setbox1=\hbox{#2}\ifdim
\wd1>\wd0\kern.5\wd1\kern-.5\wd0\fi
\copy0\kern-.5\wd0\kern-.5\wd1\copy1\ifdim\wd0>\wd1
\kern.5\wd0\kern-.5\wd1\fi}}
\def\centerover#1#2{\centeron{#1}{\setbox0=\hbox{#1}\setbox
1=\hbox{#2}\raise\ht0\hbox{\raise\dp1\hbox{\copy1}}}}
\def\centerunder#1#2{\centeron{#1}{\setbox0=\hbox{#1}\setbox
1=\hbox{#2}\lower\dp0\hbox{\lower\ht1\hbox{\copy1}}}}
\def\lsim{\;\centeron{\raise.35ex\hbox{$<$}}{\lower.65ex\hbox
{$\sim$}}\;}
\def\gsim{\;\centeron{\raise.35ex\hbox{$>$}}{\lower.65ex\hbox
{$\sim$}}\;}
\def\super#1{\ifmmode \hbox{\textsuper{#1}}\else\textsuper{#1}\fi}
\def\textsuper#1{\newcount\holdspacefactor\holdspacefactor=\spacefactor
$^{#1}$\spacefactor=\holdspacefactor}
\def\getcite#1,{\advance\citenumber by1
\ifnum\citenumber=1
\ref{#1}\let\next=\getcite\else\ifx#1@\let\next=\relax
\else ,\ref{#1}\let\next=\getcite\fi\fi\next}
\def\upon #1/#2 {{\textstyle{#1\over #2}}}
\relax

\def\mainhead#1{\setcounter{equation}{0}\addtocounter{section}{1}
  \vbox{\begin{center}\large\bf #1\end{center}}\nobreak\par}
\sectioneq

\def\til#1{\centeron{\hbox{$#1$}}{\lower 2ex\hbox{$\char'176$}}}
\def\tild#1{\centeron{\hbox{$\,#1$}}{\lower 2.5ex\hbox{$\char'176$}}}
\def\sumtil{\centeron{\hbox{$\displaystyle\sum$}}{\lower
-1.5ex\hbox{$\widetilde{\phantom{xx}}$}}}

\def\pom{{\rm P\kern -0.53em\llap I\,}}
\def\spom{{\rm P\kern -0.36em\llap \small I\,}}
\def\sspom{{\rm P\kern -0.33em\llap \footnotesize I\,}}

\newcommand{\bit}{\begin{itemize}}
\newcommand{\eit}{\end{itemize}}

\newcommand{\beq}{\begin{equation}}
\newcommand{\eeq}{\end{equation}}
\newcommand{\beqa}{\begin{eqnarray}}
\newcommand{\eeqa}{\end{eqnarray}}

\begin{document} 
\begin{titlepage} 

\rightline{\vbox{\halign{&#\hfil\cr
&ANL-HEP-PR-96-75\cr
&UF-IFT-HEP-96-18\cr
&\today\cr}}} 
\vspace{0.25in} 

\begin{center} 
 
{\large\bf 
NLO CONFORMAL SYMMETRY IN THE REGGE LIMIT OF QCD}\footnote{Work 
supported by the U.S.
Department of Energy, Division of High Energy Physics, \newline Contracts
W-31-109-ENG-38 and DEFG05-86-ER-40272} 
\medskip

{Claudio Corian\`{o},$^{b}$\footnote{
coriano@phys.ufl.edu  ~$^{\#}$arw@hep.anl.gov ~$^{+}$wusthoff@hep.anl.gov}
Alan. R. White$^{a\#}$ and 
Mark W\"usthoff $^{a+}$\ }

\vskip 0.6cm

\centerline{$^a$High Energy Physics Division}
\centerline{Argonne National Laboratory}
\centerline{9700 South Cass, Il 60439, USA.}
\vspace{0.5cm}

\centerline{$^b$Institute for Fundamental Theory}
\centerline{Department of Physics}
\centerline{ University of Florida at Gainesville, FL 32611, USA}
\vspace{0.5cm}

\end{center}

\begin{abstract}

We show that a scale invariant approximation to the next-to-leading order
BFKL kernel, constructed via transverse momentum diagrams, has a simple
conformally invariant representation in impact parameter space i.e. 
$$
\tilde{K}
(\rho_1,\rho_2,\rho_{1'},\rho_{2'})~~~=~~~g^4~N^2~ln^4\Biggl[ {
|\rho_1 - \rho_{1'}|~|\rho_2 - \rho_{2'}| \over 
|\rho_1 - \rho_{2'}|~|\rho_2 - \rho_{1'}|} \Biggr]
$$ 
That a conformally invariant representation exists is shown first by
relating the kernel directly to Feynman diagrams contributing to
two photon diffractive dissociation.

\end{abstract}

\end{titlepage}

%%%%%%%%%%%%%%%%%%%%%%%%%%%%%%%%%%%%%%%%%%%%%%%%%%%%%%%%%%

\mainhead{1. INTRODUCTION }

In the leading-log approximation, the Regge limit of QCD is described by the
BFKL equation\cite{bfkl} and the resulting ``BFKL Pomeron''. It is
well-known that the forward limit ($q\to 0$) of the BFKL equation describes
the small-x evolution of parton distributions. An important property of the
full equation is that, in impact parameter space, it is invariant under
special conformal transformations. In next-to-leading-order (NLO) this
property is expected to be lost as scale dependence enters the equation. The
direct evaluation of NLO contributions to the BFKL equation is now close to
completion\cite{FL}. 

It is nevertheless attractive to suppose that conformal symmetry could play
a vital role in solving the full dynamical problem of the QCD Pomeron. It 
would be encouraging, if this is to be the case, to see that key properties of
higher-order contributions are related to conformally symmetric 
interactions. Indeed it was conjectured in \cite{cw} that this could be the 
role of $t$-channel unitarity in determining higher-order contributions.
The results of this paper are, perhaps, a first step in the right
direction. That is we show explicit conformal symmetry for an interaction,
derived first by $t$-channel arguments, that we anticipate to be an
infra-red approximation to the NLO kernel of the BFKL equation. 

Both the leading-order BFKL kernel and the NLO approximation (which we refer 
to as the $O(g^4)$ kernel to indicate that no scale-dependence appears in 
the gauge coupling)  have been derived by a reggeon diagram technique based 
only indirectly on $t$-channel unitarity\cite{ker}. More recently it has
been shown\cite{cw} that, in part at least, these results can be obtained by
a direct analysis of the t-channel unitarity equations, analytically
continued in the complex $j$-plane. The same analysis, however, also shows
that the overall normalization of the $O(g^4)$ kernel depends on a rapidity
scale (cut-off) which cannot be fixed directly from the analysis. A similar
observation was made by Kirschner\cite{rk} who has discussed how the
$O(g^4)$ kernel emerges as an approximation when non-leading results are
obtained using the leading-order $s$-channel multi-Regge effective
action. It remains to directly connect this kernel to the explicit NLO
calculations and so determine the missing normalization factor. 

The $O(g^4)$ kernel is expressed in terms of two-dimensional
transverse momentum integrals that are directly scale invariant in
transverse momentum space. At leading-order this scale invariance property
leads to conformal invariance in impact parameter space. The main result of 
this paper is to show that the $O(g^4)$ kernel actually has a remarkably simple
representation in impact parameter space which is manifestly conformally
symmetric. The kernel $\tilde{K}$ connects two initial points
$(\rho_1,\rho_2,)$ to two final points $(\rho_{1'},\rho_{2'})$ and has the 
representation 
$$
\tilde{K}
(\rho_1,\rho_2,\rho_{1'},\rho_{2'})~~~~=~~~~g^4~N^2~ln^4[R]
\auto\label{rep}
$$ 
where 
$$
R~=~{|\rho_1 - \rho_{1'}|~|\rho_2 - \rho_{2'}| \over 
|\rho_1 - \rho_{2'}|~|\rho_2 - \rho_{1'}|}
\auto\label{RD}
$$

The above representation was actually found by realizing, as we will show 
below, that the same $O(g^4)$ kernel can be found
in a rather different context. In double diffractive dissociation of two
virtual photons the color zero exchange is modeled by two t-channel gluons
which after taking the square of the amplitude become four gluons. In this 
situation it is known how to impose conformal symmetry and so obtain a 
conformally symmetric interaction. The important result is that the
diagrams with four gluons can be rearranged in a way to be identical with
those occuring in the $O(g^4)$ kernel. Here the overall normalization is
actually fixed, but the whole contribution is of next-to-next-to-leading
order (NNLO) and does not interfere with NLO BFKL-corrections. 

Previously it had been shown\cite{cw1} that, in momentum space, the 
forward $O(g^4)$ kernel splits naturally into two components. A part 
proportional to the square of the $O(g^2)$ BFKL kernel, and a new component that
has an eigenvalue spectrum sharing many properties of the leading-order
spectrum, in particular holomorphic factorization. Holomorphic
factorization is closely related to conformal symmetry and also the square
of the $O(g^2)$ kernel has conformal properties. Therefore it was anticipated 
that the complete $O(g^4)$ kernel could have some conformal symmetry
property. The simplicity of (\ref{rep}) is nevertheless surprising.

The outline of the paper is as follows. We begin, in Section 2, by recalling 
the form of the $O(g^4)$ kernel in momentum space. In Section 3 we discuss 
the construction of the kernel from Feynman diagrams contributing to virtual 
photon-photon scattering. This construction automatically gives a 
conformally invariant result, the interesting feature is that the previously 
determined $O(g^4)$ kernel emerges. Section 4 is a brief discussion of how
Ward identity properties and other features of the BFKL equation translate
into impact parameter space. The direct evaluation of the $O(g^4)$ kernel
in impact parameter space is carried out in Section 5 and the above result
obtained. In Section 6 we discuss operators corresponding to general powers
of $R$ and suggest obtaining the spectra of such operators from a generating
function. We also give an intriguing formal representation of the BFKL
kernel involving $ln^3{R}$ and $ln^4{R}$. We conclude with a brief summary
of the results of the paper.

\mainhead{2. THE $O(g^4)$ KERNEL IN MOMENTUM SPACE}

It will be helpful for our discussion to introduce transverse momentum diagrams,
which we construct using the components illustrated in Fig.~2.1. 

\begin{center}
  \leavevmode
\begin{picture}(0,0)%
\epsfig{file=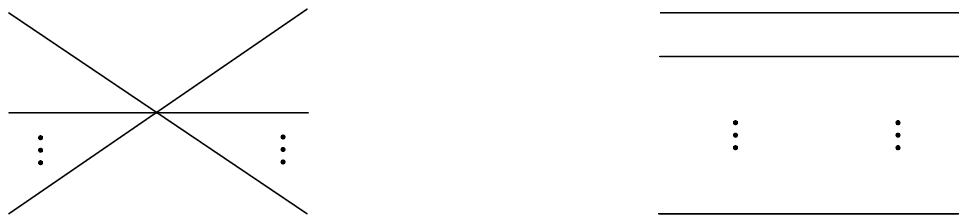}%
\end{picture}%
\setlength{\unitlength}{0.00033300in}%
\begingroup\makeatletter\ifx\SetFigFont\undefined
% extract first six characters in \fmtname
\def\x#1#2#3#4#5#6#7\relax{\def\x{#1#2#3#4#5#6}}%
\expandafter\x\fmtname xxxxxx\relax \def\y{splain}%
\ifx\x\y   % LaTeX or SliTeX?
\gdef\SetFigFont#1#2#3{%
  \ifnum #1<17\tiny\else \ifnum #1<20\small\else
  \ifnum #1<24\normalsize\else \ifnum #1<29\large\else
  \ifnum #1<34\Large\else \ifnum #1<41\LARGE\else
     \huge\fi\fi\fi\fi\fi\fi
  \csname #3\endcsname}%
\else
\gdef\SetFigFont#1#2#3{\begingroup
  \count@#1\relax \ifnum 25<\count@\count@25\fi
  \def\x{\endgroup\@setsize\SetFigFont{#2pt}}%
  \expandafter\x
    \csname \romannumeral\the\count@ pt\expandafter\endcsname
    \csname @\romannumeral\the\count@ pt\endcsname
  \csname #3\endcsname}%
\fi
\fi\endgroup
\begin{picture}(12312,4389)(916,-8235)
\put(916,-6489){\makebox(0,0)[lb]{\smash{\SetFigFont{10}{12.0}{rm}$k_n$}}}
\put(916,-4014){\makebox(0,0)[lb]{\smash{\SetFigFont{10}{12.0}{rm}$k_1$}}}
\put(916,-5214){\makebox(0,0)[lb]{\smash{\SetFigFont{10}{12.0}{rm}$k_2$}}}
\put(5626,-6489){\makebox(0,0)[lb]{\smash{\SetFigFont{10}{12.0}{rm}$k'_n$}}}
\put(3331,-8184){\makebox(0,0)[lb]{\smash{\SetFigFont{12}{14.4}{rm}(a)}}}
\put(11146,-8139){\makebox(0,0)[lb]{\smash{\SetFigFont{12}{14.4}{rm}(b)}}}
\put(8416,-6444){\makebox(0,0)[lb]{\smash{\SetFigFont{10}{12.0}{rm}$k_n$}}}
\put(8461,-4599){\makebox(0,0)[lb]{\smash{\SetFigFont{10}{12.0}{rm}$k_2$}}}
\put(8476,-4014){\makebox(0,0)[lb]{\smash{\SetFigFont{10}{12.0}{rm}$k_1$}}}
\put(5611,-4014){\makebox(0,0)[lb]{\smash{\SetFigFont{10}{12.0}{rm}$k'_1$}}}
\put(5626,-5184){\makebox(0,0)[lb]{\smash{\SetFigFont{10}{12.0}{rm}$k'_2$}}}
\end{picture}
%\begin{center}
%\leavevmode
%\epsfxsize=3.5in
%\epsffile{ke1.ps}
\vspace{0.5cm}

Fig.~2.1 (a)vertices and (b) intermediate states in transverse momentum.
\end{center}
The rules for writing amplitudes corresponding to the diagrams are the following

\begin{itemize}

\item{For each vertex, illustrated in Fig.~2.1(a), we write a factor
$$
16\pi^3\delta^2(\sum k_i~  - \sum k_i')(\sum k_i~)^2
$$}
\item{For each intermediate state, illustrated in Fig.~2.1(b), we write a factor
$$
\Gamma^k_n~=~(16\pi^3)^{-n}\int d^2k_1...d^2k_n~ /~k_1^2...k_n^2
$$}
\end{itemize}

We denote dimensionless amplitudes with a hat and remove the hat to denote the 
corresponding amplitude with the momentum conservation $\delta$-function
removed, e.g. for a 2-2 kernel $K_{2,2}$
$$
\hat{K}_{2,2}(k_1,k_2,k_{1'},k_{2'})~=~
16\pi^3\delta^2(k_1+k_2-k_{1'}-k_4) K_{2,2}(k_1,k_2,k_{1'},k_{2'})~
$$
The dimensionless amplitudes are formally scale-invariant but, in general, are 
infra-red divergent. The cancellation of such divergences is an essential 
pre-requisite for defining a scale-invariant kernel. 

In this paper we will be concerned with kernels that are symmetrized 
(or antisymmetrized) with respect to initial and final momenta. The 
diagrammatic representation of $\hat{K}^{(2)}_{2,2}$, the
$O(g^2)$ symmetrized non-forward BFKL kernel, is as shown in Fig.~2.2. 
\begin{center}
  \leavevmode
\begin{picture}(0,0)%
\epsfig{file=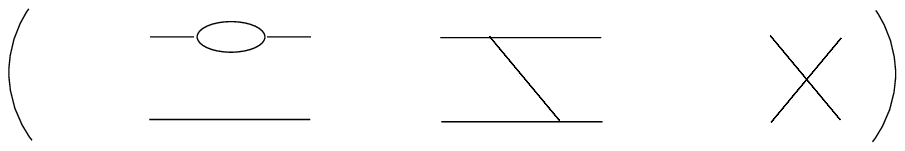}%
\end{picture}%
\setlength{\unitlength}{0.00029200in}%
\begingroup\makeatletter\ifx\SetFigFont\undefined
% extract first six characters in \fmtname
\def\x#1#2#3#4#5#6#7\relax{\def\x{#1#2#3#4#5#6}}%
\expandafter\x\fmtname xxxxxx\relax \def\y{splain}%
\ifx\x\y   % LaTeX or SliTeX?
\gdef\SetFigFont#1#2#3{%
  \ifnum #1<17\tiny\else \ifnum #1<20\small\else
  \ifnum #1<24\normalsize\else \ifnum #1<29\large\else
  \ifnum #1<34\Large\else \ifnum #1<41\LARGE\else
     \huge\fi\fi\fi\fi\fi\fi
  \csname #3\endcsname}%
\else
\gdef\SetFigFont#1#2#3{\begingroup
  \count@#1\relax \ifnum 25<\count@\count@25\fi
  \def\x{\endgroup\@setsize\SetFigFont{#2pt}}%
  \expandafter\x
    \csname \romannumeral\the\count@ pt\expandafter\endcsname
    \csname @\romannumeral\the\count@ pt\endcsname
  \csname #3\endcsname}%
\fi
\fi\endgroup
\begin{picture}(13333,1851)(5872,-4438)
\put(11901,-3687){\makebox(0,0)[lb]{\smash{\SetFigFont{17}{20.4}{rm}+ }}}
\put(15823,-3779){\makebox(0,0)[lb]{\smash{\SetFigFont{17}{20.4}{rm}$-  \; \frac{1}{2}$}}}
\put(7329,-3736){\makebox(0,0)[lb]{\smash{\SetFigFont{17}{20.4}{rm}$-\;\frac{1}{2}$}}}
\put(5872,-3736){\makebox(0,0)[lb]{\smash{\SetFigFont{17}{20.4}{rm}$\sum$}}}
\end{picture}
%\begin{center}
%\leavevmode
%\epsfxsize=4in
%\epsffile{k2.ps}
\vspace{0.2cm}

Fig.~2.2 Diagrammatic representation of $\hat{K}^{(2)}_{2,2}$.
\end{center}
The summation sign implies a sum over permutations of both the initial
and final momenta. 

The diagrammatic representation of the symmetric $O(g^4)$ 
kernel $K^{(4)}_{2,2}$ (obtained
by considering the contribution of 4-particle nonsense states to the 
$t$-channel unitarity equations) is shown in Fig.~2.3. (The values of $a$ 
and $b$ will be discussed shortly.)

\begin{center}
  \leavevmode
\begin{picture}(0,0)%
\epsfig{file=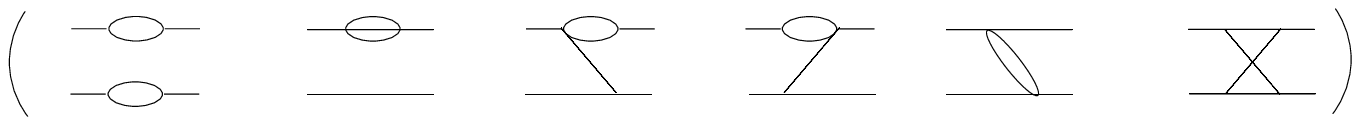}%
\end{picture}%
\setlength{\unitlength}{0.00023300in}%
\begingroup\makeatletter\ifx\SetFigFont\undefined
% extract first six characters in \fmtname
\def\x#1#2#3#4#5#6#7\relax{\def\x{#1#2#3#4#5#6}}%
\expandafter\x\fmtname xxxxxx\relax \def\y{splain}%
\ifx\x\y   % LaTeX or SliTeX?
\gdef\SetFigFont#1#2#3{%
  \ifnum #1<17\tiny\else \ifnum #1<20\small\else
  \ifnum #1<24\normalsize\else \ifnum #1<29\large\else
  \ifnum #1<34\Large\else \ifnum #1<41\LARGE\else
     \huge\fi\fi\fi\fi\fi\fi
  \csname #3\endcsname}%
\else
\gdef\SetFigFont#1#2#3{\begingroup
  \count@#1\relax \ifnum 25<\count@\count@25\fi
  \def\x{\endgroup\@setsize\SetFigFont{#2pt}}%
  \expandafter\x
    \csname \romannumeral\the\count@ pt\expandafter\endcsname
    \csname @\romannumeral\the\count@ pt\endcsname
  \csname #3\endcsname}%
\fi
\fi\endgroup
\begin{picture}(24788,1877)(1202,-4481)
\put(1202,-3779){\makebox(0,0)[lb]{\smash{\SetFigFont{14}{16.8}{rm}$\sum \;\frac{1}{2}$}}}
\put(3511,-3734){\makebox(0,0)[lb]{\smash{\SetFigFont{11}{13.2}{rm}a}}}
\put(6399,-3759){\makebox(0,0)[lb]{\smash{\SetFigFont{14}{16.8}{rm}+}}}
\put(7354,-3742){\makebox(0,0)[lb]{\smash{\SetFigFont{11}{13.2}{rm}b}}}
\put(18309,-3669){\makebox(0,0)[lb]{\smash{\SetFigFont{14}{16.8}{rm}+}}}
\put(10659,-3712){\makebox(0,0)[lb]{\smash{\SetFigFont{14}{16.8}{rm}$-$}}}
\put(14261,-3734){\makebox(0,0)[lb]{\smash{\SetFigFont{14}{16.8}{rm}$-$}}}
\put(21815,-3732){\makebox(0,0)[lb]{\smash{\SetFigFont{14}{16.8}{rm}$-  \frac{1}{2}$}}}
\end{picture}
%\begin{center}
%\leavevmode
%\epsfxsize=5in
%\epsffile{ckhat.ps}
\vspace{0.2cm}

Fig.~2.3 The diagrammatic representation of ${\hat{K}}^{4}_{2,2}$.
\end{center}
Removing the momentum conservation $\delta$-function, we have
$$
\eqalign{{1 \over (g^2N)^2} K^{(4)}_{2,2}(k_1&,k_2,k_{1'},k_{2'})
{}~=~K^{(4)}_0~+~K^{(4)}_1~+~K^{(4)}_2~+~K^{(4)}_3~+K^{(4)}_4~}.
\auto\label{sum}
$$
with
$$
\eqalign{K^{(4)}_0~=~
{a \over 2}
\sum ~ k_1^4k_2^4J_1(k_1^2)J_1(k_2^2)(16\pi^3)\delta^2(k_2-k_{2'})~,}
\auto\label{e0}
$$

$$
\eqalign{K^{(4)}_1~=~{b \over 2}~
\sum ~ k_1^4J_2(k_1^2)k_2^2(16\pi^3)\delta^2(k_2-k_{2'})~,}
\auto\label{e1}
$$

$$
\eqalign{K^{(4)}_2~=~- {1 \over 2} \sum
\Biggl({k_1^2J_1(k_1^2)k_2^2k_{2'}^2+
k_1^2k_{2'}^2J_1(k_{1'}^2)k_{1'}^2 \over
(k_1-k_{1'})^2} \Biggr),}
\auto\label{e2}
$$

$$
\eqalign{K^{(4)}_3~= ~{ 1 \over 2} \sum~
k_2^2k_{1'}^2J_1((k_1-k_{1'})^2)}
\auto\label{e3}
$$
and
$$
\eqalign{K^{(4)}_4~=~{1 \over 4}~\sum~
k_1^2k_2^2k_{1'}^2k_{2'}^2~I(k_1,k_2,k_{1'},k_{2'}), }
\auto\label{e4}
$$
where
$$
J_1(k^2)~=~{1 \over 16\pi^3}\int {d^2k' \over
(k')^2(k'-k)^2~~~,}
\auto\label{j1}
$$

$$
\eqalign{J_2(k^2)~=~{1 \over 16\pi^3}\int d^2q {1 \over
(k-q)^2}J_1(q^2)}
\auto
$$
and
$$
\eqalign{ I(k_1,k_2,k_{1'},k_{2'})~=~{1 \over 16\pi^3}\int d^2p {1 \over
p^2(p+k_1)^2(p+k_1-k_{1'})^2(p+k_{2'})^2}~~~.}
\auto\label{box}
$$
The $\sum$ again implies that we sum over permutations of
both the initial and the final state. 

In previous papers $K^{(4)}_0$ and $K^{(4)}_1$ have been defined by the 
values\cite{ker}
$$
a~~=~~0,~~~~b~~=~~{2 \over 3}
\auto\label{ab} 
$$
and also the values\cite{cw}

$$
a~~=~~1,~~~~b~~=~~{-2 \over 3}  ~.
\auto\label{ab1} 
$$
In \cite{cw1} it was incorrectly argued that the latter values are determined 
uniquely by the cancellation, after integration, of all
divergences of the complete kernel. Each of $K^{(4)}_0$ and $K^{(4)}_1$ 
contains single and double poles in $\epsilon$ when dimensionally
regularized and, at first sight, requiring the cancellation of both
singularities, after integration of the full kernel, fixes $a$ and $b$.
However, in \cite{cw1} it is shown that 
\beqa 
&&{3 \over 4}|k^2 J_1(k)|^2 ~~=~~
k^2J_2(k^2) \\
&&= {{6 {{{\rm \gamma}}^2} {{\pi }^2}}} - 
   {{{{\pi }^4}}\over {2 }} + {{12 {{\pi }^2}}\over {{{{\it \epsilon}}^2} }} + 
   {{12 {\rm \gamma} {{\pi }^2}}\over {{\it \epsilon} }} + 
   {{12 {\rm \gamma} {{\pi }^2} \log (\pi )}} + 
   {{12 {{\pi }^2} \log (\pi )}\over {{\it \epsilon} }} + 
   {{6 {{\pi }^2} {{\log (\pi )}^2}}} \nonumber \\
&& + 
   {{12 {\rm \gamma} {{\pi }^2} \log (k^2)}} + 
   {{12 {{\pi }^2} \log (k^2)}\over {{\it \epsilon} }} 
+ 
   {{12 {{\pi }^2} \log (\pi ) \log (k^2)}} + 
   {{6 {{\pi }^2} {{\log (k^2)}^2}}}.\nonumber \\
\eeqa
and so the single and double poles occur with the same relation between the
coefficients in both $K^{(4)}_0$ and $K^{(4)}_1$. As a result $a$ and $b$ 
are not determined uniquely by infra-red cancellations and it is sufficient 
only that
$$
a~+~{3b \over 4}~~=~~{1 \over 2}
\auto\label{ab2}
$$
which is, of course, satisfied by both (\ref{ab}) and (\ref{ab1}). Below we
shall find that conformal invariance determines uniquely that 
$$
a~=~{1 \over 4}~, ~~~~~b~=~{1 \over 3}~.
\auto\label{ab3}
$$

\mainhead{3. A FEYNMAN DIAGRAM CONSTRUCTION} 

We now show that the $O(g^4)$ kernel can be identified with a certain set of
Feynman diagrams as, for example, occur in the calculation of virtual
photon-photon scattering with a large rapidity gap between the fragments. As
illustrated in Fig.~3.1, (the square of the amplitude is
shown), the two photons with virtuality $Q^2$ and $Q'^2$ 
dissociate into two quark-antiquark pairs which then interact via a color-zero 
exchange represented at lowest order by two gluons. It should be pointed out
that the  diagram in Fig.~3.1 is of next-to-next-to-leading order compared
to the leading order BFKL-calculation and does not contribute at
next-to-leading-order level as the $O(g^4)$ kernel in its original form is
meant to do. Nevertheless, as we discuss, it is understood how to introduce 
conformal symmetry directly in this context. The occurence of the same
structure in two different contributions has to be taken as pure
coincidence, perhaps related to the general role of conformal symmetry. 
\begin{center}
  \leavevmode
\begin{picture}(0,0)%
\epsfig{file=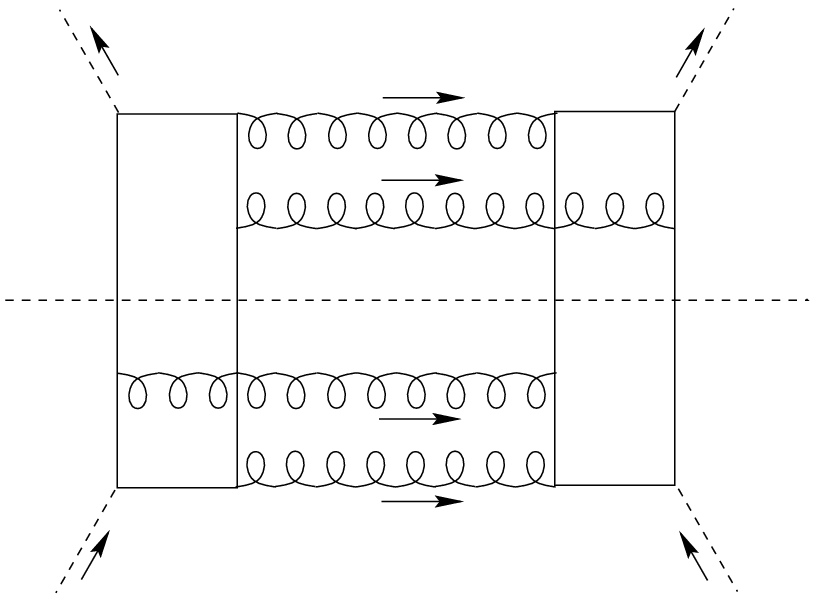}%
\end{picture}%
\setlength{\unitlength}{0.00033300in}%
\begingroup\makeatletter\ifx\SetFigFont\undefined
% extract first six characters in \fmtname
\def\x#1#2#3#4#5#6#7\relax{\def\x{#1#2#3#4#5#6}}%
\expandafter\x\fmtname xxxxxx\relax \def\y{splain}%
\ifx\x\y   % LaTeX or SliTeX?
\gdef\SetFigFont#1#2#3{%
  \ifnum #1<17\tiny\else \ifnum #1<20\small\else
  \ifnum #1<24\normalsize\else \ifnum #1<29\large\else
  \ifnum #1<34\Large\else \ifnum #1<41\LARGE\else
     \huge\fi\fi\fi\fi\fi\fi
  \csname #3\endcsname}%
\else
\gdef\SetFigFont#1#2#3{\begingroup
  \count@#1\relax \ifnum 25<\count@\count@25\fi
  \def\x{\endgroup\@setsize\SetFigFont{#2pt}}%
  \expandafter\x
    \csname \romannumeral\the\count@ pt\expandafter\endcsname
    \csname @\romannumeral\the\count@ pt\endcsname
  \csname #3\endcsname}%
\fi
\fi\endgroup
\begin{picture}(9674,7053)(1084,-7447)
\put(9556,-1203){\makebox(0,0)[lb]{\smash{\SetFigFont{9}{10.8}{rm}$Q'$}}}
\put(5659,-3532){\makebox(0,0)[lb]{\smash{\SetFigFont{9}{10.8}{rm}$k_2$}}}
\put(5688,-1213){\makebox(0,0)[lb]{\smash{\SetFigFont{9}{10.8}{rm}$k_1$}}}
\put(5632,-6974){\makebox(0,0)[lb]{\smash{\SetFigFont{9}{10.8}{rm}$k_4$}}}
\put(5676,-4587){\makebox(0,0)[lb]{\smash{\SetFigFont{9}{10.8}{rm}$k_3$}}}
\put(9604,-6879){\makebox(0,0)[lb]{\smash{\SetFigFont{9}{10.8}{rm}$Q'$}}}
\put(1540,-1163){\makebox(0,0)[lb]{\smash{\SetFigFont{9}{10.8}{rm}$Q$}}}
\put(1461,-7034){\makebox(0,0)[lb]{\smash{\SetFigFont{9}{10.8}{rm}$Q$}}}
\end{picture}
\vspace{0.5cm}

Fig.~3.1 Scattering of two virtual photons with rapidity gap.
\end{center}

Following ref.\cite{BarWu} the coupling of four gluons to the quark-loop
can be summarized in seven terms:
\newpage
\beqa\label{MWe1}
&&D_{(4,0)}^{(1;+,+)}(k_1,k_2,k_3,Q^2)\;=\;g^2 N \;A
\cdot\nonumber\\
&&\left\{\;D_{(2;0)}(k_1,Q^2)\,+\,D_{(2;0)}(k_2,Q^2)\,
+\,D_{(2;0)}(k_3,Q^2)\,+\,D_{(2;0)}(k_1+k_2+k_3,Q^2)\right.\nonumber
\\
&&\;\;\left.-\,D_{(2;0)}(k_1+k_2,Q^2)\,-\,D_{(2;0)}(k_1+k_3,Q^2)\,
-\,D_{(2;0)}(k_2+k_3,Q^2)\right\}
\eeqa
with an overall constant $A$ which is not of relevance in this
discussion. Although not needed explicitly, for completeness
we give the two functional forms for $D_{(2;0)}$  (depending on the
polarization of the photon):
\beqa\label{MWe2}
D^t_{(2;0)}(k,Q^2)&=&\sum_f e_f^2\alpha_s\,\frac{\sqrt{N^2-1}}{2\pi}
\;\int_0^1 d\alpha
\int_0^1dy\frac{[1-2\alpha(1-\alpha)][1-2y(1-y)]\,k^2}{y(1-y)k^2+
\alpha(1-\alpha)Q^2}\nonumber\\
\\
D^l_{(2;0)}(k,Q^2)&=&\sum_f e_f^2\alpha_s\,\frac{\sqrt{N^2-1}}{2\pi}
\;\int_0^1 d\alpha
\int_0^1dy\frac{[2\alpha(1-\alpha)][2y(1-y)]\,k^2}{y(1-y)k^2+
\alpha(1-\alpha)Q^2}\;\;.
\eeqa

The index (1;+,+) stands for the color projection and symmetry of the
gluon pair above and below the central cut in Fig.~3.1.
In this example `1' means color singlet and `+' means even under interchange of
the two gluons. A very important property of $D^{(1;+,+)}_{(4;0)}$ is the fact
that it vanishes whenever one of the momenta $k_1,...,k_4$ becomes zero.
This property, which we will refer to as color cancellation,
still holds in the case of an even color octet state.
The odd color octet configuration, however, fails to provide complete color
cancellation and diagrams like that in Fig.~3.1 with odd signature are not
infrared safe. The infrared singularities are supposed to cancel with those
from contributions at next-to-next-to-leading order real gluon emission. 

In our case where we required color singlet exchange,
no infrared problems occur. The singularities due to the gluon
propagators for $k_i=0$ are cancelled. We get the $O(g^4)$ kernel by
taking ``the square" of expression (\ref{MWe1}), i.e. by adding the
propagators and integrating over the transverse phase space:
\beqa\label{MWe3}
&&\frac{1}{(16\pi^3)^3}\;\int \frac{d^2k_1}{k_1^2}
\frac{d^2k_2}{k_2^2}\frac{ d^2k_3}{k_3^2}\frac{ d^2k_4}{k_4^2}
\;\delta^2(k_1+k_2+k_3+k_4)\nonumber\\ 
&&D^{(1;+,+)}_{(4;0)}(k_1,k_2,k_3,Q^2)
D^{(1;+,+)}_{(4,0)}(k_1,k_2,k_3,Q'^2) \\
&&~=~\frac{A^2}{(16\pi^3)^2}
\;\int \frac{d^2k}{k^2}\frac{d^2k'}{{k'}^2} \;
D_{(2,0)}(k,Q^2)D_{(2,0)}(k',Q'^2)\;\frac{1}{24} K^{(4)}_{2,2}(k,-k,k',-k')\;\;.
\nonumber\eeqa
The expansion of the lhs of eq.(\ref{MWe3}) is straightforward. A quick
check can be done by adding the absolute values of the coefficient in
eqs. (\ref{e0})-(\ref{e4}) and multiplying the 
result by 24 (note that (\ref{e2}) contains two diagrams and
should be counted twice). Provided that the coefficient $a$ is equal to 1/4
and the coefficient $b$ equal to 1/3 one finds (\ref{sum}) agrees directly
with the 7x7 terms of the lhs of eq.(\ref{MWe3}). 

After having established the relation between the $O(g^4)$ kernel
and the diagram in Fig.~3.1, we know from ref.\cite{lip2} 
that due to color cancellation the gluon propagators in impact
parameter space may be rewritten in a conformally invariant way.
(We illustrate this for two gluon exchange in the next Section).
This implys that a conformally invariant representation of
the $O(g^4)$ kernel necessarily exists. It remains only to find the precise 
form. This we do via the explicit calculations below.
 
\mainhead{4. STRUCTURE OF THE INTEGRAL EQUATION} 

We first write the BFKL equation in momentum space in terms of the
Green's function $f$:
$$
\eqalign{
&f(\omega, k_1,k_2,k_{1'},k_{2'})\;=\;f^0(\omega,k_1,k_2,k_{1'},k_{1'})\cr
&\hspace{1cm}+\;{1 \over \omega}~~{1 \over 16\pi^3} 
\int {d^2k_{1''} \over k_{1''}^2} {d^2k_{2''} \over k_{2''}^2} 
~\hat{K}^{(2)}_{2,2}(k_1,k_2,k_{1''},k_{2''})
f(\omega, k_{1''},k_{2''},k_{1'},k_{2'})}
\auto\label{mom}
$$
where, as we have discussed in Section 2, 
$\hat{K}^{(2)}_{2,2}(k_1,k_2,k_{1'},k_{2'})$
is a dimensionless kernel including a momentum conserving
$\delta$-function. 
The inhomogeneous term $f^0$ in (\ref{mom}) is given by  
the two gluon propagator, i.e. the starting configuration for a gluon-ladder:
$$ 
f^0(\omega,k_1,k_2,k_{1'},k_{2'})~=~{1 \over \omega }
~{\delta^2(k_1-k_{1'}) \over k_1^2}
{\delta^2(k_2-k_{2'}) \over k_2^2}\;\;.
\auto\label{gll}
$$
In the complete scattering amplitude $f^0$ is sandwiched between two colorless
states described by ``wave functions" $F$ and $F'$ with the following
``Ward-identity" property:
$$
F(k_1,k_2)~
\centerunder{$\longrightarrow$} {\raisebox{-5mm} {$k_i \to 0 $ }} 0
~~~~i~=1,2
$$
and
$$
F'(k_{1'},k_{2'})~
\centerunder{$\longrightarrow$} {\raisebox{-5mm} {$k_{i'} \to 0 $ }} 0
~~~~{i'}~=1',2'
\auto\label{wd}
$$
This property is due to color cancellation and ensures an infrared-stable 
result.

The corresponding form of the BFKL equation in impact parameter space
is then 
$$
\eqalign{
&\tilde{f}(\omega, \rho_1,\rho_2,\rho_{1'},\rho_{2'})\;=\;
~\tilde{f}^0(\omega,\rho_1,\rho_2,\rho_{1'},\rho_{2'})\cr
&\hspace{0.5cm}+\;{1 \over \omega}
 ~\int d\rho_{1''}d\rho_{1''}^* d\rho_{2''}d\rho_{2''}^*~
\tilde{K}^{(2)}_{2,2}(\rho_1,\rho_2,\rho_{1''},\rho_{2''})
|\partial_{1''}|^2|\partial_{2''}|^2
\tilde{f}(\omega, \rho_{1''},\rho_{2''},\rho_{1'},\rho_{2'})}
\auto\label{im}
$$
($\partial_{1''} = \partial / \partial \rho_{1''}$ etc.) with
a conformally invariant kernel
$\tilde{K}^{(2)}_{2,2}(\rho_1,\rho_2,\rho_{1'},\rho_{2'})$. 
Explicit representations can be found in \cite{BarWu2,lip2}.
In this context $\tilde{K}^{(2)}_{2,2}(\rho_1,\rho_2,\rho_{1''},\rho_{2''})$
is simply defined in a formal way to be the fourier 
transform of the ``reduced'' kernel
$$
\hat{K}^{(2)}_{2,2}(k_1,k_2,k_{1''},k_{2''}) /
k_1^2 k_2^2 k^2_{1''} k_{2''}^2
\auto\label{red}
$$
The momentum factors had to be added to compensate for the dimensional 
dependence introduced by the fourier transform.

The ``Ward-identity" property (\ref{wd})
translates into impact parameter space as
$$
\int d^2 \rho_i \tilde{F}(\rho_1,\rho_2)~=~0~~~~i~=1,2~
$$
and
$$
\int d^2 \rho_{i'} \tilde{F'}(\rho_{1'},\rho_{2'})~=~0~~~~{i'}~=1',2'~.
\auto\label{wd1}
$$
In order to write down the inhomogeneous term $\tilde{f^0}$ in impact
parameter space we need to know the Fourier transform of a simple
propagator. We use the formula
$$
\int d^2 k { e^{i k.\rho} \over (k^2 + m^2)} ~= ~K_0(m|\rho|)
\auto\label{bes}
$$
where $K_0$ is the modified Bessel function, and take the limit
$m \rightarrow 0$:
$$
K_0(m|\rho|) ~~\centerunder{$\longrightarrow$} {\raisebox{-5mm} {$m \to 0 $ }} 
~~~ - ln[m|\rho|/2] ~+~\psi(1)~+~O(m)~.
\auto\label{bes1}
$$
By means of eq.(\ref{wd1}) the $\ln[m/2]$ terms drop out and we are left
with:
$$
\tilde{f^0}(\omega,\rho_1,\rho_2,\rho_1',\rho_2')~=~
{4 \over (2\pi )^4\omega} \ln|\rho_{11'}|\ln|\rho_{22'}|
\auto\label{f0}
$$
with the notation $\rho_1-\rho_{1'}=\rho_{11'}$. 

One can now directly
show, again with property (\ref{wd1}), that the rhs of eq.(\ref{gll}) is
conformally invariant. Equivalently we can simply simply exploit (\ref{wd1}
to add extra terms which lead to an explicitly conformally invariant 
expression \cite{lip2} (this is what we will do for the four gluon kernel
in the next Section). The result is  
$$
\eqalign{\tilde{f^0}(\omega,\rho_1,\rho_2,\rho_1',\rho_2')~&=~
{4 \over (2\pi )^4\omega} \ln^2 \left| \frac{\rho_{11'}\rho_{22'}}
{\rho_{12'}\rho_{1'2}}\right|\cr
&=~{4 \over (2\pi )^4\omega} \ln^2 ~R~}
\auto\label{gll1}
$$
which differs from (\ref{f0}) only by terms which give zero contribution by 
virtue of (\ref{wd1}). $R$ is the harmonic ratio previously defined in
(\ref{RD}). Note that we have defined the Green's function to be symmetric
under interchange of 1 and 2 or 1' and 2'. 

\mainhead{5. $O(g^4)$ KERNEL IN IMPACT PARAMETER SPACE}

We now discuss the fourier transforms of each term in $\hat{K}^{(4)}_{2,2}$. 
We shall need just the one basic transform, i.e. (\ref{bes}) of the previous
section.

We begin with the most complicated case i.e. the box diagram $K^{(4)}_4$. We 
use the notation of Fig.~4.1

\begin{center}
  \leavevmode
\begin{picture}(0,0)%
\epsfig{file=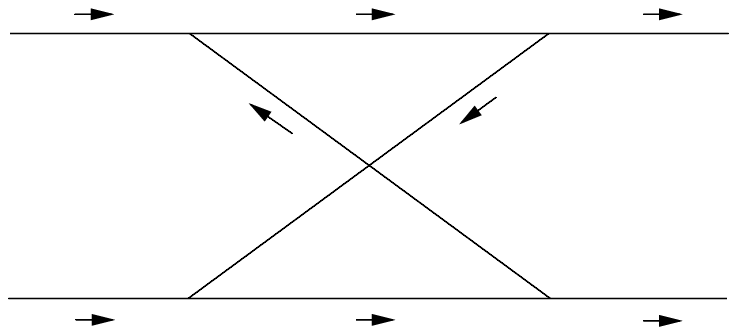}%
\end{picture}%
\setlength{\unitlength}{0.00033300in}%
\begingroup\makeatletter\ifx\SetFigFont\undefined
% extract first six characters in \fmtname
\def\x#1#2#3#4#5#6#7\relax{\def\x{#1#2#3#4#5#6}}%
\expandafter\x\fmtname xxxxxx\relax \def\y{splain}%
\ifx\x\y   % LaTeX or SliTeX?
\gdef\SetFigFont#1#2#3{%
  \ifnum #1<17\tiny\else \ifnum #1<20\small\else
  \ifnum #1<24\normalsize\else \ifnum #1<29\large\else
  \ifnum #1<34\Large\else \ifnum #1<41\LARGE\else
     \huge\fi\fi\fi\fi\fi\fi
  \csname #3\endcsname}%
\else
\gdef\SetFigFont#1#2#3{\begingroup
  \count@#1\relax \ifnum 25<\count@\count@25\fi
  \def\x{\endgroup\@setsize\SetFigFont{#2pt}}%
  \expandafter\x
    \csname \romannumeral\the\count@ pt\expandafter\endcsname
    \csname @\romannumeral\the\count@ pt\endcsname
  \csname #3\endcsname}%
\fi
\fi\endgroup
\begin{picture}(8680,5100)(2124,-7980)
\put(9795,-3324){\makebox(0,0)[lb]{\smash{\SetFigFont{12}{14.4}{rm}$k'_1$}}}
\put(6285,-3264){\makebox(0,0)[lb]{\smash{\SetFigFont{12}{14.4}{rm}$p+k_1$}}}
\put(3000,-3264){\makebox(0,0)[lb]{\smash{\SetFigFont{12}{14.4}{rm}$k_1$}}}
\put(4665,-5259){\makebox(0,0)[lb]{\smash{\SetFigFont{12}{14.4}{rm}$p$}}}
\put(2985,-7884){\makebox(0,0)[lb]{\smash{\SetFigFont{12}{14.4}{rm}$k_2$}}}
\put(6270,-7884){\makebox(0,0)[lb]{\smash{\SetFigFont{12}{14.4}{rm}$p+k'_2$}}}
\put(9780,-7884){\makebox(0,0)[lb]{\smash{\SetFigFont{12}{14.4}{rm}$k'_2$}}}
\put(7950,-5154){\makebox(0,0)[lb]{\smash{\SetFigFont{12}{14.4}{rm}$p+k_1-k'_1$}}}
\end{picture}
%\begin{center}
%\leavevmode
%\epsfxsize=2.5in
%\epsffile{boxcr.ps}
\vspace{0.5cm}

Fig.~4.1 Notation for the crossed box diagram.
\end{center}
and consider the transform of ``reduced'' diagram as discussed in the
previous Section  i.e. 
$$ 
\eqalign{ \tilde{K}^{(4)}_4(\rho_1,\rho_2,\rho_{1'},\rho_{2'}) 
~=~{1 \over 4} \sum \int& d^2k_1d^2k_2d^2k_{1'}d^2k_{2'}\cr
&e^{ik_1.\rho_1 + ik_2.\rho_2 - ik_{1'}.\rho_{1'} - ik_{2'}.\rho_{2'}}\cr
&\delta^2(k_1 + k_2 - k_{1'} -k_{2'}) I(k_1, k_2, k_{1'}, k_{2'}) }
\auto\label{fou}
$$

where once again summation is over permutation of the initial and final 
momenta. (In this case, since $I(k_1, k_2, k_{1'}, k_{2'})$ is completely 
symmetric, the factor of $1/4$ simply cancels the effect of the summation.) 
Using the $\delta$-function to perform the $k_{2'}$ integration and 
inserting (\ref{box}) we obtain, using (\ref{bes}) and (\ref{bes1}),

$$
\eqalign{ \tilde{K}^{(4)}_4(\rho_1,\rho_2,\rho_{1'},\rho_{2'})
~=&~\int d^2k_1d^2k_2d^2k_{1'}d^2p
~~e^{-i(\rho_1 - \rho_{2'})p - i(\rho_{1'} - \rho_1)(p +k_1)}\cr
&~~~~\times~ e^{-i(\rho_2 - \rho_{1'})(p -k_{1'} +k_1) 
-i(\rho_{2'} - \rho_2)(p -k_{1'} +k_1 +k_2)}\cr
&~~~~\times~{1 \over
p^2(p+k_1)^2(p+k_1-k_{1'})^2(p+-k_{1'} +k_1 +k_2)^2}\cr
=&~[- ln|\rho_1 - \rho_{2'}|~-~ln|m|~+~ln2~+~\psi(1)~+~O(m)]\cr
&~~~~\times [- ln|\rho_2 - \rho_{1'}|~-~ln|m|~+~...]\cr
&~~~~\times [- ln|\rho_1 - \rho_{1'}|~-~...]
[- ln|\rho_2 - \rho_{2'}| ~+~ ...]\cr
\equiv&~ln|\rho_{12'}|  ln|\rho_{21'}|  ln|\rho_{11'}|  
ln|\rho_{22'}|~+~....}  
\auto\label{fou1}
$$
where, in the last line, the omitted terms involve factors of $ln |m|$,
$\psi(1)$ etc.. 

Consider next the disconnected diagrams contained in $K^{(4)}_0$. 
For the reduced diagram we simply have, using the
notation of Fig.~4.2,
$$
\eqalign{ \tilde{K}^{(4)}_0(\rho_1,\rho_2,\rho_{1'},\rho_{2'}) 
~=~{a \over 2} \sum \int& d^2k_1d^2k_2d^2k_{1'}d^2k_{2'}\cr
&e^{ik_1.\rho_1 + ik_2.\rho_2 - ik_{1'}.\rho_{1'} - ik_{2'}.\rho_{2'}}\cr
&\delta^2(k_1 - k_{1'})~\delta^2(k_2 - k_{2'})~ J_1(k_1)~ J_1(k_2) }
\auto\label{k0}
$$
Using the $\delta$-functions to perform the $k_{1'}$ and $k_{2'}$ 
integrations we obtain
$$
\eqalign{ \tilde{K}^{(4)}_0(\rho_1,\rho_2,\rho_{1'},\rho_{2'})
~=&~{a \over 2} \sum \int d^2k_1d^2k_2d^2p_1d^2p_2
~~e^{i(\rho_1 - \rho_{1'})(k_1+p_1) + i(\rho_{1'} - \rho_1)p_1}\cr
&~~~~\times~ e^{i(\rho_2 - \rho_{2'})(k_2 + p_2) 
~+~i(\rho_{2'} - \rho_2)p_2}\cr
&~~~~\times~{1 \over
p_1^2p_2^2(p_1 +k_1)^2(p_1+k_2)^2}\cr
=&~{a \over 2} \sum 
[- ln|\rho_1 - \rho_{1'}|~-~ln|m|~+~ln2~+~\psi(1)~+~O(m)]^2\cr
&~~~~\times [- ln|\rho_2 - \rho_{2'}|~-~ln|m|~+~...]^2\cr
\equiv&~{a \over 2} \sum ~ln^2|\rho_{11'}|  ln^2|\rho_{22'}| ~+~....}  
\auto\label{k01}
$$
\begin{center}
\leavevmode
\begin{picture}(0,0)%
\epsfig{file=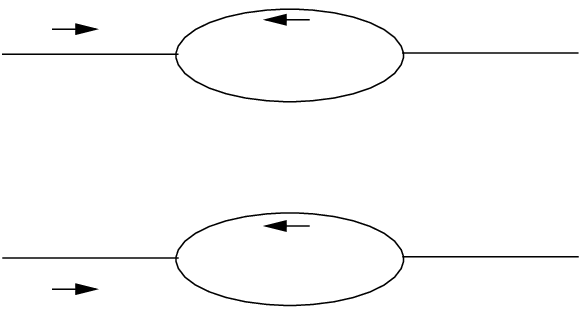}%
\end{picture}%
\setlength{\unitlength}{0.00033300in}%
\begingroup\makeatletter\ifx\SetFigFont\undefined
% extract first six characters in \fmtname
\def\x#1#2#3#4#5#6#7\relax{\def\x{#1#2#3#4#5#6}}%
\expandafter\x\fmtname xxxxxx\relax \def\y{splain}%
\ifx\x\y   % LaTeX or SliTeX?
\gdef\SetFigFont#1#2#3{%
  \ifnum #1<17\tiny\else \ifnum #1<20\small\else
  \ifnum #1<24\normalsize\else \ifnum #1<29\large\else
  \ifnum #1<34\Large\else \ifnum #1<41\LARGE\else
     \huge\fi\fi\fi\fi\fi\fi
  \csname #3\endcsname}%
\else
\gdef\SetFigFont#1#2#3{\begingroup
  \count@#1\relax \ifnum 25<\count@\count@25\fi
  \def\x{\endgroup\@setsize\SetFigFont{#2pt}}%
  \expandafter\x
    \csname \romannumeral\the\count@ pt\expandafter\endcsname
    \csname @\romannumeral\the\count@ pt\endcsname
  \csname #3\endcsname}%
\fi
\fi\endgroup
\begin{picture}(6960,4485)(2964,-6999)
\put(3496,-2874){\makebox(0,0)[lb]{\smash{\SetFigFont{12}{14.4}{rm}$k_1$}}}
\put(6136,-3549){\makebox(0,0)[lb]{\smash{\SetFigFont{12}{14.4}{rm}$p_1$}}}
\put(6136,-6114){\makebox(0,0)[lb]{\smash{\SetFigFont{12}{14.4}{rm}$p_2$}}}
\put(3496,-6909){\makebox(0,0)[lb]{\smash{\SetFigFont{12}{14.4}{rm}$k_2$}}}
\end{picture}
%\begin{center}
%\leavevmode
%\epsfxsize=2in
%\epsffile{k0.ps}
%\vspace{0.3cm}

Fig.~4.2 Notation for the disconnected $K_0$ diagram.
\end{center}
Proceeding in the same manner we obtain
$$
\eqalign{ \tilde{K}^{(4)}_1(\rho_1,\rho_2,\rho_{1'},\rho_{2'})
~=~{b \over 2} \sum ln^3|\rho_{11'}| ln|\rho_{22'}| ~+~....}  
\auto\label{k1}
$$

$$
\eqalign{ \tilde{K}^{(4)}_2(\rho_1,\rho_2,\rho_{1'},\rho_{2'})
~=&~{-1 \over 2}  
\sum ln^2|\rho_{11'}|(ln|\rho_{12'}| ln|\rho_{22'}| ~+~.... \cr
&~+~ln^2|\rho_{11'}| ln|\rho_{21'}| ln|\rho_{22'}| ~+~....) }  
\auto\label{k2}
$$

$$
\eqalign{ \tilde{K}^{(4)}_3(\rho_1,\rho_2,\rho_{1'},\rho_{2'})
~=&~{1 \over 2} \sum ln|\rho_{11'}| ln^2|\rho_{12'}| ln|\rho_{22'}| ~+~.... }
\auto\label{k3}
$$

Next we note that $ln^4~R$ has the expansion
$$
\eqalign{ ln^4~\biggl[{|\rho_{11'}|~|\rho_{22'}| \over
|\rho_{11'}|~|\rho_{22'}|}\biggr]
~&=~(ln|\rho_{12'}| + ln|\rho_{21'}| - ln|\rho_{11'}|  
-ln|\rho_{22'}|)^4\cr   
&=~6~ln^2|\rho_{11'}|ln^2|\rho_{22'}|~+~6~ ln^2|\rho_{12'}|  
ln^2|\rho_{21'}|
~+~4 ln^3|\rho_{11'}| ln|\rho_{22'}|\cr
&~~~+~4~ln|\rho_{11'}|  
ln^3|\rho_{22'}|
~+~4 ln^3|\rho_{12'}| ln|\rho_{21'}| ~+~4~ln|\rho_{12'}|  
ln^3|\rho_{21'}|\cr
&~~~-~12 ln^2|\rho_{11'}| ln|\rho_{12'}| ln|\rho_{22'}| ~-~12~ln^2|\rho_{11'}|  
ln|\rho_{21'}| ln|\rho_{22'}|\cr 
&~~~-~12 ln|\rho_{11'}| ln|\rho_{12'}| ln^2|\rho_{22'}| ~-~12~ln^2|\rho_{11'}|  
ln|\rho_{21'}| ln|\rho_{22'}|\cr 
&~~~-~12 ln^2|\rho_{12'}| ln|\rho_{11'}| ln|\rho_{21'}| ~-~12~ln^2|\rho_{12'}|  
ln|\rho_{22'}| ln|\rho_{21'}|\cr 
&~~~-~12 ln|\rho_{12'}| ln|\rho_{11'}| ln^2|\rho_{21'}| ~-~12~ln^2|\rho_{12'}|  
ln|\rho_{22'}| ln|\rho_{21'}|\cr 
&~~~+~12 ln|\rho_{11'}| ln^2|\rho_{12'}| ln|\rho_{22'}| ~+~12~ln|\rho_{11'}|  
ln^2|\rho_{21'}| ln|\rho_{22'}|\cr 
&~~~+~12 ln|\rho_{12'}| ln^2|\rho_{11'}| ln|\rho_{21'}| ~+~12~ln|\rho_{12'}|  
ln^2|\rho_{22'}| ln|\rho_{21'}|\cr 
&~~~+~24(ln|\rho_{12'}|  ln|\rho_{21'}|  ln|\rho_{11'}|  
ln|\rho_{22'}|)~+~....}
\auto\label{ln4}
$$
where the omitted terms are all independent of one (or more) of   
$\rho_1$, $\rho_2$, $\rho_{1'}$ and $\rho_{2'}$. 

We can rewrite (\ref{ln4}) as 
$$
\eqalign{ {1 \over 24} 
&ln^4~\biggl[{|\rho_{11'}|~|\rho_{22'}| \over
|\rho_{11'}|~|\rho_{22'}|}\biggr]
=~{1\over 8} \sum ~ln^2|\rho_{11'}|ln^2|\rho_{22'}| 
~+~{1\over 6} \sum ~ ln^3|\rho_{12'}| ln|\rho_{21'}| \cr
&~~-~{1\over 2} \sum  (ln^2|\rho_{11'}| ln|\rho_{12'}| ln|\rho_{22'}| 
~-~ln^2|\rho_{11'}|  ln|\rho_{21'}| ln|\rho_{22'}|)\cr 
&~~+~{1\over 2} \sum ~ ln|\rho_{11'}| ln^2|\rho_{12'}| ln|\rho_{22'}|\cr
&~~+~{1\over 4} \sum (ln|\rho_{12'}|  ln|\rho_{21'}|  ln|\rho_{11'}|  
ln|\rho_{22'}|)
~+~....}
\auto\label{ln41}
$$
From (\ref{fou1}) - (\ref{k3}), it is then clear that, with the values of $a$ 
and $b$ given by (\ref{ab3}), (\ref{ln4}) gives the
fourier transform of the complete sum of terms in $\hat{K}^{(4)}_{2,2}$,
apart from an overall multiplicative factor and apart from terms that are
independent of one (or more) of $\rho_1$, $\rho_2$, $\rho_{1'}$ and
$\rho_{2'}$, and, or, involve factors of $ln |m|$, $ln2$ and $\psi(1)$. 
In fact the infra-red cancellation of the factors of $ln [m]$ in 
$\hat{K}^{(4)}_{2,2}$ implies that the factors of $ln2$ and $\psi(1)$
must also cancel. The terms that are
independent of one of $\rho_1$, $\rho_2$, $\rho_{1'}$ and $\rho_{2'}$
can be dropped by virtue of (\ref{wd1}) and the following
discussion. It follows that (\ref{rep}) can be used directly as the fourier
transform $\tilde{K}^{(4)}_{2,2}$ of $\hat{K}^{(4)}_{2,2}$ in the BFKL
equation (\ref{im}). 

\mainhead{6. HIGHER-ORDER GENERALISATIONS AND SPECTRA}

It is interesting to note that the distinctive terms in 
(\ref{ln41}) each corresponds to one of the diagrams of Fig.~5.1, which are 
impact parameter analogs of the transverse momentum diagrams originally used 
to define the kernel. Each line in these diagrams corresponds to a logarithm
(propagator) of the relative impact parameter of the two points joined. We 
have shown above that the sum of all diagrams in which two pairs of points are
joined by four propagators is associated with the conformally invariant
operator $ln^4R$. We have also seen in Section 3 that the sum of diagrams
in which the two pairs are joined by two propagators (i.e. the two gluon
propagator) is associated with $ln^2 R$. 

\begin{center}
  \leavevmode
\begin{picture}(0,0)%
\epsfig{file=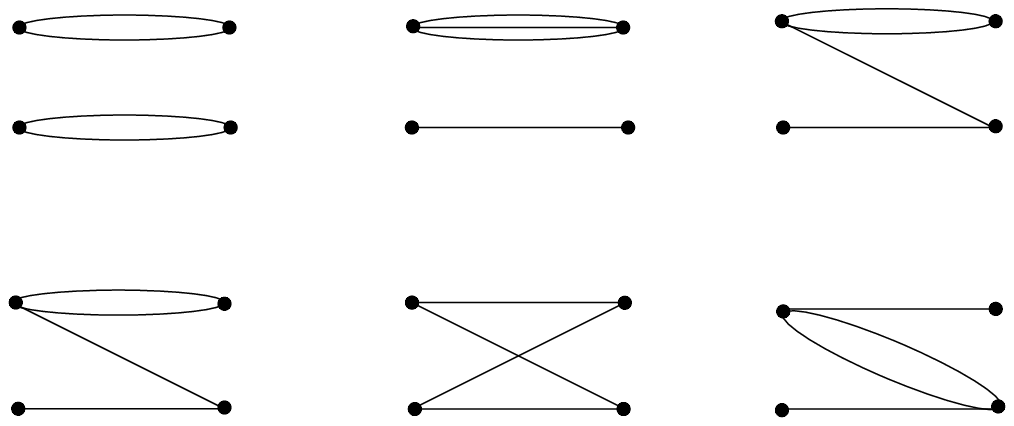}%
\end{picture}%
\setlength{\unitlength}{0.00033300in}%
\begingroup\makeatletter\ifx\SetFigFont\undefined
% extract first six characters in \fmtname
\def\x#1#2#3#4#5#6#7\relax{\def\x{#1#2#3#4#5#6}}%
\expandafter\x\fmtname xxxxxx\relax \def\y{splain}%
\ifx\x\y   % LaTeX or SliTeX?
\gdef\SetFigFont#1#2#3{%
  \ifnum #1<17\tiny\else \ifnum #1<20\small\else
  \ifnum #1<24\normalsize\else \ifnum #1<29\large\else
  \ifnum #1<34\Large\else \ifnum #1<41\LARGE\else
     \huge\fi\fi\fi\fi\fi\fi
  \csname #3\endcsname}%
\else
\gdef\SetFigFont#1#2#3{\begingroup
  \count@#1\relax \ifnum 25<\count@\count@25\fi
  \def\x{\endgroup\@setsize\SetFigFont{#2pt}}%
  \expandafter\x
    \csname \romannumeral\the\count@ pt\expandafter\endcsname
    \csname @\romannumeral\the\count@ pt\endcsname
  \csname #3\endcsname}%
\fi
\fi\endgroup
\begin{picture}(12347,4921)(631,-5568)
\put(646,-991){\makebox(0,0)[lb]{\smash{\SetFigFont{12}{14.4}{rm}1}}}
\put(631,-2176){\makebox(0,0)[lb]{\smash{\SetFigFont{12}{14.4}{rm}2}}}
\put(4111,-991){\makebox(0,0)[lb]{\smash{\SetFigFont{12}{14.4}{rm}1'}}}
\put(4096,-2176){\makebox(0,0)[lb]{\smash{\SetFigFont{12}{14.4}{rm}2'}}}
\end{picture}
%\begin{center}
%\leavevmode
%\epsfxsize=4in
%\epsffile{impa.ps}
\vspace{0.5cm}

Fig.~5.1 Impact parameter diagrams. 
\end{center}

As a generalization, we can expect that for arbitrary $m$ we can associate
$ln^mR$ with diagrams in which the two pairs of points are joined by $m$
propagators. It is natural to conjecture that such interactions will appear
at the appropriate order in the BFKL kernel in some form of conformal
approximation. This conjecture was made for the corresponding transverse
momentum diagrams in \cite{uni}. We should note, however, that it is far 
from clear that the
appropriate $t$-channel unitarity construction can actually be carried
through. It is possible that the explicit relationship of the $O(g^4)$
kernel to the exact NLO calculations could provide a better understanding of
whether such higher-order interactions should be expected to appear. 

At first sight it might be thought that $ln^3R$ is directly related to
the BFKL kernel. However, the antisymmetry under $1'\leftrightarrow 2'$
implies that only the ``odd'' part of the kernel is obtained. Similarly the
$ln^2~R$ and $ln^4~ R$ kernels are necessarily symmetric. In general the
kernels are odd (even) under $1'\leftrightarrow  2'$ when $m$ is odd(even). 

The eigenvalue spectrum of the kernels we are discussing are defined with
respect to eigenfunctions labeled by  $(\nu,n)$, $-\infty <\nu < \infty, ~n
= 0, \pm 1, \pm 2, ...$ (in momentum space, for $q=0$, the eigenfunctions are
$\phi_{\nu,n}= |k|^{\nu}e^{i{n \over 2}\theta}$). Symmetry or antisymmetry
under $1' \leftrightarrow 2'$ 
determines $n$ to be even or odd respectively. In \cite{cw1} it was shown 
that the spectrum of $K^{(4)}_{2,2}$ (which is even in $n$) can be written
in the form 
$$ 
{\cal E}(\nu,n)~=~{1 \over \pi} [\chi(\nu,n)]^2~-~\Lambda(\nu,n) ~.
\auto\label{opr}
$$
where $\chi(\nu,n)$ are the eigenvalues of the (even part of) the BFKL 
kernel and 
$$
\eqalign{ \Lambda(\nu,n) ~=~-~{1 \over 4\pi}
\biggl(\beta'\bigl({|n| + 1\over 2} + 
i\nu\bigr)
~+~\beta'\bigl({|n| + 1 \over 2} -i\nu\bigr)\biggr). }
\auto\label{lam}
$$
$\beta(x)$ is the incomplete beta function, i.e.
$$
\eqalign{
\beta(x)~&=~\int^1_0 dy~y^{x -1}[1+y]^{-1} }
\auto
$$

The $\Lambda(\nu,n)$ are the eigenvalues of a separately infra-red finite
component ${\cal K}_2$ of $K^{(4)}_{2,2}$ extracted from the box diagram
$K^{(4)}_4$. The $\Lambda(\nu,n)$ also have a holomorphic factorisation
property. From the above discussion, and from (\ref{opr}) in particular, it
appears that the BFKL kernel $K_{BFKL}$ has the formal 
representation 
$$
K_{BFKL}~=~c_1~ln^3~R ~+~c_2~[ln^4~R ~-~{\cal K}_2]^{1\over 2}~,  
\auto\label{def}
$$
where $c_1$ and $c_2$ can easily be calculated. $ln^3~R$ provides the 
antisymmetric part of the kernel and $[ln^4~R 
~-~{\cal K}_2]^{1\over 2}$ the symmetric part. The representation of the
symmetric part, however, is non-trivial and needs more investigation.
The hope is that with ${\cal K}_2$ suitably defined in impact 
parameter space one finds a simple form similar to the antisymmetric
part.

It is interesting that all the kernels $ln^m~R$ can be simultaneously 
studied by considering derivatives (with repect to $\delta$) of the 
generating function
$$
\cal{G}(R,\delta)~=~R^{\delta}
\auto\label{gen}
$$
It appears that a number of interesting spectra and relationships between 
operators, including (\ref{def}), can be derived from this starting point.
This subject is currently under study and will be discussed in a future
publication. 

\section{\bf Conclusions}

By finding the surprisingly simple representation (\ref{rep}), we  
have shown explicitly the conformal invariance of a 
contribution to the NLO BFKL kernel constructed via t-channel unitarity.
This lends some support to the conjecture\cite {cw1} that $t$-channel
unitarity determines conformally invariant interactions underlying
non-leading log contributions to the Regge limit of QCD. We have also 
generated a further set of candidate interactions, i.e. 
$$
ln^m~R~\equiv~
ln^m\Biggl[ {
|\rho_1 - \rho_{1'}|~|\rho_2 - \rho_{2'}| \over 
|\rho_1 - \rho_{2'}|~|\rho_2 - \rho_{1'}|} \Biggr]
\auto\label{repm}
$$ 
and suggested a method for their study.

It is important that the representation (\ref{rep}) was found 
via a Feynman diagram construction that leads to the same $O(g^4)$ kernel.
The diagrams involved do not contribute at the level of perturbation
theory associated with the t-channel construction. Nevertheless the 
Feynman diagram construction may prove useful in understanding the 
general significance of higher-order interactions such as (\ref{repm}). 

\newpage

\centerline{\bf Acknowledgements}
C.C. thanks the Theory Group at Argonne and in particular L. E. Gordon 
for hospitality.

\end{document}